\documentclass[longauth]{aa} 
\usepackage{graphicx}
\usepackage{txfonts}
\usepackage{lscape}
\usepackage{multirow}
\usepackage[authoryear]{natbib}
\begin{document}
    \renewcommand{\topfraction}{0.85}
    \renewcommand{\bottomfraction}{0.7}
    \renewcommand{\textfraction}{0.15}
    \renewcommand{\floatpagefraction}{0.66}

\title{H.E.S.S. observations of $\gamma$-ray bursts in 2003--2007}

\small{
\author{F. Aharonian\inst{1,13}
 \and A.G.~Akhperjanian \inst{2}
 \and U.~Barres de Almeida \inst{8} \thanks{supported by CAPES Foundation, Ministry of Education of Brazil}
 \and A.R.~Bazer-Bachi \inst{3}
 \and B.~Behera \inst{14}
 \and W.~Benbow \inst{1}
 \and K.~Bernl\"ohr \inst{1,5}
 \and C.~Boisson \inst{6}
 \and A.~Bochow \inst{1}
 \and V.~Borrel \inst{3}
% ...
}
}

\offprints{P.H. Tam, \email{ phtam@lsw.uni-heidelberg.de }}

\institute{
Max-Planck-Institut f\"ur Kernphysik, P.O. Box 103980, D 69029
Heidelberg, Germany
\and
 Yerevan Physics Institute, 2 Alikhanian Brothers St., 375036 Yerevan,
Armenia
\and
Centre d'Etude Spatiale des Rayonnements, CNRS/UPS, 9 av. du Colonel Roche, BP
4346, F-31029 Toulouse Cedex 4, France
\and
Universit\"at Hamburg, Institut f\"ur Experimentalphysik, Luruper Chaussee
149, D 22761 Hamburg, Germany
\and
Institut f\"ur Physik, Humboldt-Universit\"at zu Berlin, Newtonstr. 15,
D 12489 Berlin, Germany
\and
LUTH, Observatoire de Paris, CNRS, Universit\'e Paris Diderot, 5 Place Jules Janssen, 92190 Meudon,
France
Obserwatorium Astronomiczne, Uniwersytet Ja
\and
IRFU/DSM/CEA, CE Saclay, F-91191
Gif-sur-Yvette, Cedex, France
\and
University of Durham, Department of Physics, South Road, Durham DH1 3LE,
U.K.
\and
Unit for Space Physics, North-West University, Potchefstroom 2520,
    South Africa
\and
Laboratoire Leprince-Ringuet, Ecole Polytechnique, CNRS/IN2P3,
 F-91128 Palaiseau, France
\and
Laboratoire d'Annecy-le-Vieux de Physique des Particules, CNRS/IN2P3,
9 Chemin de Bellevue - BP 110 F-74941 Annecy-le-Vieux Cedex, France
\and
Astroparticule et Cosmologie (APC), CNRS, Universite Paris 7 Denis Diderot,
10, rue Alice Domon et Leonie Duquet, F-75205 Paris Cedex 13, France
\thanks{UMR 7164 (CNRS, Universit\'e Paris VII, CEA, Observatoire de Paris)}
\and
Dublin Institute for Advanced Studies, 5 Merrion Square, Dublin 2,
Ireland
\and
Landessternwarte, Universit\"at Heidelberg, K\"onigstuhl, D 69117 Heidelberg, Germany
% ...
}

\abstract
   {}
   {Very-high-energy (VHE; $\ga$100~GeV) $\gamma$-rays are expected from
     $\gamma$-ray bursts (GRBs) in some scenarios. Exploring this photon energy regime is
     necessary for understanding the energetics and properties of GRBs. }
   {GRBs have been one of the prime targets for the H.E.S.S. experiment, which
     makes use of four Imaging Atmospheric Cherenkov Telescopes
     (IACTs) to detect VHE $\gamma$-rays. Dedicated observations of 32
     GRB positions were made in the years 2003--2007 and a search for VHE
     $\gamma$-ray counterparts of these GRBs was made. Depending on the
     visibility and observing conditions, the observations mostly start
     minutes to hours after the burst and typically last two hours.}
   {Results from observations of 22
     GRB positions are presented and evidence of a VHE signal was found neither in observations of any individual
     GRBs, nor from stacking data from subsets of GRBs with higher expected VHE
     flux according to a model-independent ranking scheme. Upper
     limits for the VHE $\gamma$-ray flux from the GRB
     positions were derived. For those GRBs with measured redshifts,
     differential upper limits at the energy threshold after correcting for absorption due to
     extra-galactic background light are also presented.}
   {}

   \maketitle

\section{Introduction}

Gamma-ray bursts (GRBs) are the most energetic events in
the $\gamma$-ray regime. Depending on their duration (e.g. $T_{90}$), GRBs are categorized into long GRBs ($T_{90}>2$~s) and short GRBs ($T_{90}<2$~s). First detected in late 1960s~\citep{klebe73}, GRBs remained mysterious for three decades.
Breakthroughs in understanding GRBs came only after
the discovery of longer-wavelength afterglows
with the launch of \emph{BeppoSAX} in 1997~\citep{paradijs00}. Multi-wavelength (MWL) observations have proved to be crucial in our understanding of GRBs, and provide valuable information about their physical
properties. These MWL afterglow observations are generally explained by synchrotron emission from shocked electrons in the relativistic \emph{fireball} model~\citep{piran99,zhang04}. A plateau phase is revealed in many of the \emph{Swift}/XRT light curves, the origin of which is still not clear~\citep{zhang06}. Observations of GRBs at energies $>$10~GeV may test some of the ideas that have been suggested to explain the X-ray observations~\citep{fan08}.

In the framework of the relativistic \emph{fireball} model, photons with energies up to $\sim$10
TeV or higher are expected from the GRB afterglow phase \citep{zhang04,fan08b}. Possible leptonic radiation mechanisms include forward-shocked electrons up-scattering self-emitted synchrotron
photons~\citep[SSC processes;][]{dermer00,zhang01,fan08} or photons from other shocked
regions~\citep{wang01}. Physical parameters, such as the ambient density of the surrounding
material ($n$), magnetic field equipartition fraction ($\epsilon_B$), and bulk
Lorentz factor ($\Gamma_{\rm bulk}$) of the outflow, may be constrained by observations at these energies
\citep{wang01,peer05}.

A possible additional contribution to VHE emission
relates to the X-ray flare phenomenon. X-ray flares are
found in more than 50\% of the \emph{Swift} GRBs during the afterglow phase \citep{chincarini07}. The
energy fluence of some of them (e.g. GRB~050502B) is comparable to that of the
prompt emission. Most of them
are clustered at $\sim$10$^2$--10$^3$s after the GRB~\citep[see Figure~2
  in][]{chincarini07}, while late X-ray flares ($>$10$^4$s) are also
observed; when these happen they can cause an increase in the X-ray flux of
an order of magnitude or more over the power-law temporal
decay~\citep{curran08}. The cause of X-ray flares is still a subject of
debate, but corresponding VHE $\gamma$-ray flares from inverse-Compton (IC) processes
are predicted \citep{wang06,galli07,fan08}. The accompanying external-Compton flare may be
weak if the flare originated behind the
external shock, e.g. from prolonged central engine activity~\citep{fan08}. However, in the
external shock model, the expected SSC flare is very strong at GeV energies
and can be readily detected using a VHE instrument
with an energy threshold of $\sim$100 GeV~\citep{galli08}, such as
the H.E.S.S. array, for a typical GRB at z$\sim$1. Therefore, VHE $\gamma$-ray data taken during an X-ray flare may help for distinguishing the internal/external shock origin of the X-ray flares, and may be used as a diagnostic tool for the late central engine
activity.

\citet{waxman00} and \citet{Murase08} suggest that GRBs may be sources of ultra-high-energy cosmic rays (UHECRs). In this case, $\pi$-decays from proton-$\gamma$ interaction may generate VHE emission. The VHE
$\gamma$-ray emission produced from such a hadronic component is generally expected to
decay more slowly than the leptonic sub-MeV radiation \citep{Boettcher98}. \citet{dermer07}
suggests a combined leptonic/hadronic scenario to explain the
rapidly-decaying phase and plateau phase seen in many of the \emph{Swift}/XRT
light curves. This model can be tested with VHE observations taken minutes to hours after the burst.

Most searches for VHE $\gamma$-rays from GRBs have obtained negative results
\citep{connaughton97,atkins05}. There may be indications of excess photon
events from some observations, but these results are not conclusive
\citep{amenomori96,padilla98,atkins00,poirier03}. Currently, the most
sensitive detectors in the VHE $\gamma$-ray regime are IACTs. \citet{horan07} presented upper limits from 7 GRBs observed with the Whipple Telescope during the pre-\emph{Swift} era. Upper limits for 9 GRBs with redshifts that were either unknown or $>$3.5 were also reported by the MAGIC collaboration
\citep{albert07}. In general,
these limits do not violate a power-law extrapolation of the keV spectra
obtained with satellite-based instruments. However, most GRBs are now believed to originate at cosmological distances, therefore absorption of VHE $\gamma$-rays by the EBL \citep{nikishov62} must be considered when interpreting these limits. 

In this paper, observations of 22 $\gamma$-ray bursts made with
H.E.S.S. during the years 2003--2007 are reported. They represent the largest sample of GRB afterglow observations made by an IACT array and result in the most stringent upper limits obtained in the VHE band. The prompt phase of GRB~060602B was observed serendipitously with H.E.S.S. The results of
observations before, during, and after this burst are presented in~\citet{aha09}.


\section{The H.E.S.S. experiment and GRB observation strategy}

The H.E.S.S. array\footnote{http://www.mpi-hd.mpg.de/hfm/HESS/HESS.html} is a
system of four 13m-diameter IACTs located at 1\,800 m above sea level in the
Khomas Highland of Namibia ($23\degr16\arcmin18\arcsec$~S,
$16\degr30\arcmin00\arcsec$~E). Each of the four telescopes is located at a
corner of a square with a side length of 120 m. This configuration was
optimized for maximum sensitivity to $\sim$100~GeV photons. The effective
collection area increases from $\sim$10$^3\mathrm{m}^2$ at 100 GeV to more than
$10^5\mathrm{m}^2$ at 1 TeV for observations at a zenith angle (Z.A.) of 20$\degr$. The
system has a point source sensitivity above 100 GeV of $\sim$1.4$\times$10$^{-11}
\mathrm{erg}\,\mathrm{cm}^{-2}\,\mathrm{s}^{-1}$ ($3.5\%$ of the flux from the
Crab nebula) for a $5 \sigma $ detection in a 2~h observation. Each
H.E.S.S. camera consists of 960 photomultiplier tubes (PMTs), which in total
provide a field of view (FoV) of $\sim$5$\degr$. This relatively large
FoV allows for the simultaneous determination of the background events from
off-source positions, so that no dedicated off run is
needed~\citep{aha06c}. The slew rate of the array is $\sim$100$\degr$ per
minute, enabling it to point to any sky position within $\sim$2 minutes. The
H.E.S.S. array is currently the only IACT array in the Southern Hemisphere
used for an active GRB observing
programme\footnote{http://www.lsw.uni-heidelberg.de/projects/hess/HESS/grbs.phtml}.

The trigger system of the H.E.S.S. array is described in~\citet{funk04}. The
stereoscopic technique is used, i.e. a coincidence of at least two telescopes
triggering within a window of (normally) 80 nanoseconds is required. This
largely rejects background events caused by local muons that trigger only a single
telescope.

The observations reported here were obtained over the period March 2003 to
October 2007.
The observations of two GRBs in 2003 were made using two telescopes while the system was
under construction. Before 2003 July, each of the two telescopes took data
separately. Stereo analysis was then performed on the data, which requires
coincidence of events to be determined offline using GPS time stamps. After the installation of the
central trigger system in 2003 July, the stereo multiplicity requirement
was determined on-line. All observations since 2004
have made use of the completed four-telescope array and the stereo technique~\citep{aha06c}.

Most of the data were taken in 28 minute runs using \emph{wobble} mode, i.e. the
GRB position is placed at an offset, $\theta_\mathrm{offset}$, of $\pm0\fdg5$ or $0\fdg7$ (in R.A. and Decl.) relative to the centre of the camera FoV during observations. 

Onboard GRB triggers distributed by the {\it Swift}
satellite, as well as triggers from \emph{INTEGRAL} and \emph{HETE-II} confirmed by
ground-based analysis, are followed by H.E.S.S. observations. Upon the reception of a GCN\footnote{The Gamma ray bursts Coordinates Network, http://gcn.gsfc.nasa.gov/} notice from one of
these satellites (with appropriate indications\footnote{which include, e.g., burst position incompatible with known sources, and a high signal-to-noise ratio of the burst} that the source is a genuine GRB),
the burst position is observed if Z.A.$\la$45$\degr$ (to ensure a
reasonably low-energy threshold) during H.E.S.S. dark
time\footnote{H.E.S.S. observations are taken in darkness and when the moon is
  below the horizon. The fraction of H.E.S.S. dark time is about 0.2}. An
automated program is running on site to keep the shift crew alerted of any new
detected GRBs in real time. Depending on the observational constraints and the
measured redshifts of the GRBs reported through GCN circulars\footnote{http://gcn.gsfc.nasa.gov/gcn3\_archive.html}, observations of the burst positions are started up to $\sim$24 hours after the burst time,
typically with an exposure time of $\approx$120 minutes in \emph{wobble}
mode. The remarkably nearby, bright GRB~030329 was an exceptional case. It was not observed
until 11.5 days after the burst because of poor weather, which
prohibited observation any earlier. 



\section{The GRB observations}

Thirty-two GRBs were observed with H.E.S.S. during the period from March 2003
to October 2007. After applying a set of data-quality criteria that rejects observation runs with non-optimal weather conditions and hardware status, 22 GRB observations were selected for analysis and are described in this section.


\begin{table*}
\begin{minipage}[t]{180mm}
\caption{Properties of GRBs observed with H.E.S.S. from March 2003 to October 2007.}
\label{GRBtable1}
\centering
\renewcommand{\footnoterule}{}  
\begin{tabular}{lccll@{}rc@{}c@{}rc@{}c@{}cl@{}r}
\hline\hline
      GRB   & Satellite  & Trigger & R.A.\footnote{R.A., Decl., and the positional errors (90\% containment) were taken from GCN Reports (http://gcn.gsfc.nasa.gov/report\_archive.html) for GRB~061110A -- GRB~071003 and GCN Circulars otherwise.} & Decl.$^a$ & Error$^a$ & Energy band &
      Fluence\footnote{Fluence and $T_\mathrm{90}$ data for
        GRB~050726 -- GRB~070612B were taken from~\citet{sakamoto08} except that
        $T_\mathrm{90}$ of GRB~060505 was taken
        from~\citet{Palmer06_GCN5076}. Fluence and $T_\mathrm{90}$ data of
        GRB~030329 and GRB~030821 were taken from \citet{sakamoto05}, and those of GRB~041006 from \citet{Shirasaki08}. Other data were taken from GCN Circulars and \emph{HETE} pages (http://space.mit.edu/HETE/Bursts).}
      & $T_\mathrm{90}^b$ & X\footnote{X: X-ray, O: optical, R: radio;
        ``$\surd$" indicates the detection of a counterpart, ``$\times$" a
        null detection, and ``$.$" that no measurement was reported in
        the corresponding energy range, from
        http://grad40.as.utexas.edu/grblog.php} & O$^c$ & R$^c$ & $z$ & Rank\footnote{The
        relative expected VHE flux for each GRB is ranked according to
        the empirical scheme described in Sect.~\ref{sect_rank}}  \\
            &              & number &                            &      & ($\arcsec$) & (keV) & ($10^{-8}$\,erg\,cm$^{-2}$) & (s)  &  &  & \\
    \hline
    071003  & \emph{Swift} & 292934 & $20^\mathrm{h}07^\mathrm{m}24\fs25$ & +$10\degr56\arcmin48\farcs8$  & 5.7 & 15--150     & 830     &
    $\sim$150      & $\surd$ & $\surd$ & $\surd$   & 1.604\footnote{\citet{perley08}}    & 5  \\
    070808  & \emph{Swift} & 287260 & $00^\mathrm{h}27^\mathrm{m}03\fs36$ & +$01\degr10\arcmin34\farcs8$  & 1.9 & 15--150     & 120     & $\sim$32        & $\surd$ & $\surd$ & $.$       & $\cdots$ & 9  \\
    070724A & \emph{Swift} & 285948 & $01^\mathrm{h}51^\mathrm{m}13\fs96$ & -$18\degr35\arcmin40\farcs1$  & 2.2 & 15--150     & 3       &
    $\sim$0.4    & $\surd$ & $\times$ & $\times$ & 0.457\footnote{\citet{Cucchiara07_GCN6665}}  & 21 \\
    070721B & \emph{Swift} & 285654 & $02^\mathrm{h}12^\mathrm{m}32\fs95$ & -$02\degr11\arcmin40\farcs6$  & 0.9 & 15--150     & 360     & $\sim$340      & $\surd$ & $\surd$ & $\times$  & 3.626\footnote{\citet{Malesani07_GCN6651}}  & 10 \\
    070721A & \emph{Swift} & 285653 & $00^\mathrm{h}12^\mathrm{m}39\fs24$ & -$28\degr22\arcmin00\farcs6$  & 2.3 & 15--150     & 7.1     & 3.868   & $\surd$ & $\surd$ & $.$ & $\cdots$ & 20 \\
    070621  & \emph{Swift} & 282808 & $21^\mathrm{h}35^\mathrm{m}10\fs14$ & -$24\degr49\arcmin03\farcs1$  & 2   & 15--150     & 430     & 33      & $\surd$ & $\times$ & $.$ & $\cdots$ & 1  \\
    070612B & \emph{Swift} & 282073 & $17^\mathrm{h}26^\mathrm{m}54\fs4$  & -$08\degr45\arcmin08\farcs7$  & 4.7 & 15--150     & 168     & 13.5    & $\surd$ & $\times$ & $.$ & $\cdots$ & 15 \\
    070429A & \emph{Swift} & 277571 & $19^\mathrm{h}50^\mathrm{m}48\fs8$  & -$32\degr24\arcmin17\farcs9$  & 2.4 & 15--150     & 91      & 163.3   & $\surd$ & $\surd$ & $.$ & $\cdots$ & 3  \\
    070419B & \emph{Swift} & 276212 & $21^\mathrm{h}02^\mathrm{m}49\fs57$ & -$31\degr15\arcmin49\farcs7$  & 3.5 & 15--150     & 736     & 236.4   & $\surd$ & $\surd$ & $.$ & $\cdots$ & 7  \\
    070209  & \emph{Swift} & 259803 & $03^\mathrm{h}04^\mathrm{m}50^\mathrm{s}$  & -$47\degr22\arcmin30\arcsec$  & 168 & 15--150     & 2.2     & 0.09
    & $\times$ & $\times$ & $.$     & 0.314?\footnote{Redshift of a candidate host galaxy~\citet{berger6101}.} & 22 \\
    061110A & \emph{Swift} & 238108 & $22^\mathrm{h}25^\mathrm{m}09\fs9$  & -$02\degr15\arcmin30\farcs7$  & 3.7 & 15--150     & 106     & 40.7    & $\surd$ & $\surd$ & $.$ & 0.758\footnote{\citet{Fynbo07_GCN6759}}   & 11 \\
    060526  & \emph{Swift} & 211957 & $15^\mathrm{h}31^\mathrm{m}18\fs4$  & +$00\degr17\arcmin11\farcs0$  & 6.8 & 15--150     & 126     & 298.2   & $\surd$ & $\surd$ & $.$ & 3.21\footnote{\citet{berger5170}}   & 8 \\
    060505  & \emph{Swift} & 208654 & $22^\mathrm{h}07^\mathrm{m}04\fs50$ & -$27\degr49\arcmin57\farcs8$  & 4.7 & 15--150     & 94.4    & $\sim$4 & $\surd$ & $\surd$ & $.$ & 0.0889\footnote{\citet{ofek06}} & 18 \\
    060403  & \emph{Swift} & 203755 & $18^\mathrm{h}49^\mathrm{m}21\fs80$ & +$08\degr19\arcmin45\farcs3$  & 5.5 & 15--150     & 135     & 30.1    & $\surd$ & $\times$ & $.$ & $\cdots$ & 16 \\
    050801  & \emph{Swift} & 148522 & $13^\mathrm{h}36^\mathrm{m}35^\mathrm{s}$ & -$21\degr55\arcmin41\arcsec$ & 1   & 15--150     & 31      & 19.4    & $\surd$ & $\surd$ & $\times$ & 1.56\footnote{Redshift according to~\citet{pasquale07}, based on afterglow modelling}  & 2  \\
    050726  & \emph{Swift} & 147788 & $13^\mathrm{h}20^\mathrm{m}12\fs30$ & -$32\degr03\arcmin50\farcs8$  & 6   & 15--150     & 194     & 49.9    & $\surd$ & $\surd$ & $.$ & $\cdots$ & 13 \\
    050509C & \emph{HETE-II}      & H3751  & $12^\mathrm{h}52^\mathrm{m}53\fs94$ & -$44\degr50\arcmin04\farcs1$  & 1   & 2--30       & 60      & 25      & $\surd$ & $\surd$ & $\surd$ & $\cdots$ & 19 \\
    050209  & \emph{HETE-II}      & U11568 & $08^\mathrm{h}26^\mathrm{m}$ & +$19\degr41\arcmin$ & 420 & 30--400     & 200     & 46 & $.$ & $\times$ & $.$ & $\cdots$ & 14 \\
    041211B\footnote{Although this burst was referred to as GRB~041211 in
      various GCN Circulars, the proper name GRB~041211B~\citep[e.g.,
        in][]{pelangeon06} should be used to distinguish it from another
      burst, GRB~041211A (=H3621) which occurred earlier on the same day
      (P\'elangeon, A., private communication).}  & \emph{HETE-II}      & H3622  & $06^\mathrm{h}43^\mathrm{m}12^\mathrm{s}$ & +$20\degr23\arcmin42\arcsec$ & 80  & 30--400     & 1000    & $>$100  & $.$ & $\times$ & $.$  & $\cdots$ & 4  \\
    041006  & \emph{HETE-II}      & H3570  & $00^\mathrm{h}54^\mathrm{m}50\fs23$ & +$01\degr14\arcmin04\farcs9$  & 0.1 & 30--400       & 713    & $\sim$20 & $\surd$ & $\surd$ & $\surd$ & 0.716\footnote{\citet{soderberg06}}  & 6 \\
    030821  & \emph{HETE-II}      & H2814  & $21^\mathrm{h}42^\mathrm{m}$  & -$44\degr52$   & \footnote{The position error of this burst is large, see Fig.~\ref{030821overlay}}    & 30--400   & 280     & 23      & $.$ & $.$ & $.$       & $\cdots$ & 17 \\
    030329  & \emph{HETE-II}      & H2652  & $10^\mathrm{h}44^\mathrm{m}49\fs96$ & +$21\degr31\arcmin17\farcs44$ & $10^{-3}$ & 30--400   & 10760   & 33      & $\surd$ & $\surd$ & $\surd$ & 0.1687\footnote{\citet{stanek03}} & 12 \\
    \hline
\end{tabular}
\end{minipage}
\end{table*}


\subsection{Properties of the GRBs}

For each burst, the observational properties as obtained from the triggering satellite are
shown in Table~\ref{GRBtable1}. These include trigger number, energy band,
fluence in that energy band, and the duration of the burst
($T_{90}$). Whenever there were follow-up observations in the X-ray, optical, or
radio bands, whether a detection has occurred (denoted by a tick $\surd$) or not (denoted by a cross
$\times$) is also shown. If no observation at a given wavelength was
reported, a dot~($.$) is shown. The reported redshifts ($z$) of 10 GRBs
are also presented, of which 6 are lower than one. Two observed bursts,
GRB~070209 and GRB~070724A, are short
GRBs while the rest are long GRBs. The population of short GRBs has a
redshift distribution~\citep{Berger07} significantly less than that of the long
GRBs~\citep{Jakobsson06}. Therefore, on average they are likely to suffer from a lower level of EBL absorption.

X-ray flares were detected from three of the GRBs in the H.E.S.S. sample. They occurred at 273s
after the burst for GRB~050726, 284s for GRB~050801, and
$2.6\times10^5$s for GRB~070429A~\citep{curran08}. Unfortunately, the flares occurred
outside the time windows of the H.E.S.S. observations.

\subsection{H.E.S.S. observations}

For each burst, the start time, $T_{\rm start}$, of the H.E.S.S. observations
after the burst is shown in Table~\ref{GRB_stat}. Since an observing strategy to start observing the burst position up to $\sim$24 hours after the burst time is applied,
the mean $T_{\rm start}$ is of the
order of 10 hours. The (good-quality) exposure time of the observations using
$N_{\rm tel}$ telescopes for each burst is included. The mean Z.A. of the
observations is also presented.



\subsection{The ranking scheme}
\label{sect_rank}
As mentioned in the introduction, there is no lack of models predicting VHE emission from GRBs. However, the evolution of the possible VHE $\gamma$-ray emission with time is model-dependent. To give an empirical, model-independent estimate of the relative expected
VHE flux of each GRB (which also depends on $T_{\rm start}$), it is assumed that: (1) the relative VHE signal scales
as the energy released in the prompt emission, taken as a typical energy
measure of a GRB. Hence $F_{\rm VHE} \propto F_{15-150 \rm
  \,keV}$ where $F_{15-150 \rm \,keV}$ is the fluence in the {\em Swift}/BAT band. For bursts not triggered by BAT, the measured fluence is extrapolated into this energy band; (2) the possible VHE signal fades as
time goes on, as observed in longer wavelength (e.g. X-ray) data. In particular, the VHE
flux follows the average decay of the X-ray flux and therefore $F_{\rm VHE} \propto F_{15-150 \rm \,keV}\times
t^{-1.3}$ where $t$ denotes the time after the burst and 1.3 is the average X-ray afterglow late-time power-law decay index \citep{nousek06}. Since in most cases the exposure time of
the observations is much shorter than $T_{\rm start}$ (the start time of the
corresponding H.E.S.S. observations after the trigger), the expected
flux at $T_{\rm start}$ can be used as
a measure of the strength of the VHE signal, and therefore of the relative
possibility of detecting a VHE signal from that GRB. By setting $t$ to $T_{\rm start}$, we have
\begin{equation}\label{rank_eqn}
    F_{\rm VHE} \propto F_{15-150 \rm \,keV}\times T_{\rm start}^{-1.3}
\end{equation}
The rank of each GRB according to equation~\ref{rank_eqn} is shown in the last column in Table~\ref{GRBtable1}. Note that redshift information (available for only a few GRBs), and thus the corresponding
EBL absorption, is not taken into account in the ranking scheme.


\section{Data Analysis}

Calibration of data, event reconstruction and rejection of the cosmic-ray background (i.e. $\gamma$-ray event selection criteria) were performed as described in~\citet{aha06c}, which employs the techniques described by~\citet{hillas96}.

Gamma-like events were then taken from a circular region (on-source) of radius
$\theta_{\rm cut}$ centred at the burst position given in
Table~\ref{GRBtable1}. The background was
estimated using the reflected-region background model as described
in~\citet{berge07}, in which the number of background events in the on-source
region ($N_\mathrm{off}$) is
estimated from $n_\mathrm{region}$ off-source regions located at the same $\theta_\mathrm{offset}$ as the on-source region during the same observation. The number of
$\gamma$-like events is given by $N_{\rm on}-\alpha N_{\rm off}$ where $N_{\rm
  on}$ is the total number of events detected in the on-source
region and $\alpha=1/n_\mathrm{region}$ the normalization factor.

Independent analyses of various GRBs using different methods and background
estimates~\citep{berge07} yielded consistent results.

\subsection{Analysis technique}
\label{sample_analysis}
Two sets of analysis cuts were applied
to search for a VHE $\gamma$-ray signal from observational data taken with three or four telescopes. These are `standard' cuts~\citep{aha06c} and `soft' cuts\footnote{`Soft' cuts were called `spectrum' cuts in~\citet{aha06a}.}~\citep[the latter have lower energy thresholds, as described in][]{aha06a}.
For standard (soft) cuts, $\theta_{\rm cut}=0.11\degr$ ($\theta_{\rm cut}=0.14\degr$). While standard cuts are optimized for a source with a power-law spectrum of photon index $\Gamma=2.6$, soft cuts are optimized for a source with a steep spectrum ($\Gamma=5.0$), and have better sensitivity at lower energies. Since EBL absorption is less severe for lower energy photons, the soft-cut analysis is useful in searching for VHE $\gamma$-rays from GRBs which are at cosmological distances. For example, the photon indices of two blazars PKS~2005-489 \citep{aha05} and PG~1553+113 \citep{aha08} were measured to be $\Gamma\ga4$.

An exception to this analysis scheme is GRB~030329. As the central trigger
system had yet to be installed when this observation was made, a slightly
different analysis technique was used. The description of the image and
analysis cuts used for the data from GRB~030329 can be found in \citet{aha05}.
For GRB~030821, only the standard-cut analysis (for two-telescope data) was
performed (see Sect.~\ref{sect_030821}).

The positional error circle of most GRBs, with the exceptions of GRB~030821, GRB~050209, and
GRB~070209, is small compared to the H.E.S.S. point spread function (PSF). The 68\%
  $\gamma$-ray containment radius, $\theta_{68}$, of the H.E.S.S. PSF can be as small
as $\sim$3$\arcmin$, depending on the Z.A. and $\theta_\mathrm{offset}$ of the observations, and the analysis cuts applied. The 68\% containment radius, $\theta_{68}$, of the observations of GRB~050209 and
GRB~070209 is about 9$\arcmin$ using standard-cut analysis\footnote{$\theta_{68}$ is larger using soft-cut analysis}, slightly larger
than the corresponding error circles. Therefore, point-source analyses were
performed for all GRBs except GRB~030821, the error box of which is much bigger than the H.E.S.S. PSF (see Sect.~\ref{sect_030821} for its treatment).

\subsection{Energy threshold}

The energy threshold, $E_{\rm th}$, is conventionally defined as the peak in
the differential $\gamma$-ray rate versus energy
curve of a fictitious source with photon index $\Gamma$~\citep{konopelko99}. This curve is a
convolution of the effective area with the expected energy spectrum of the
source as seen on Earth. Such energy thresholds, obtained by the
standard-cut analysis and the soft-cut analysis for each GRB observation, are
shown in Table~\ref{GRB_stat}, assuming $\Gamma=2.6$.
The energy threshold depends on the Z.A. of the observations and the analysis used.
The larger the Z.A., the higher is the energy threshold. Moreover, soft-cut
analysis gives a lower value of $E_{\rm th}$ than that of standard-cut
analysis. Note that $\gamma$-ray photons with energies below $E_{\rm th}$ can
be detected by the telescopes.

\subsection{Optical efficiency of the instrument}

The data presented were also corrected for the long-term changes in the optical
efficiency of the instrument. The optical efficiency has decreased over a period
of a few years. This has changed the effective area and energy
threshold of the instrument. Specifically, the energy threshold has increased
with time. Using images of local muons in the FoV, this effect in the calculation of flux upper limits is corrected~\citep[c.f.][]{aha06c}.

\section{Results}

\begin{figure}
   \centering
   \includegraphics[width=9.5cm]{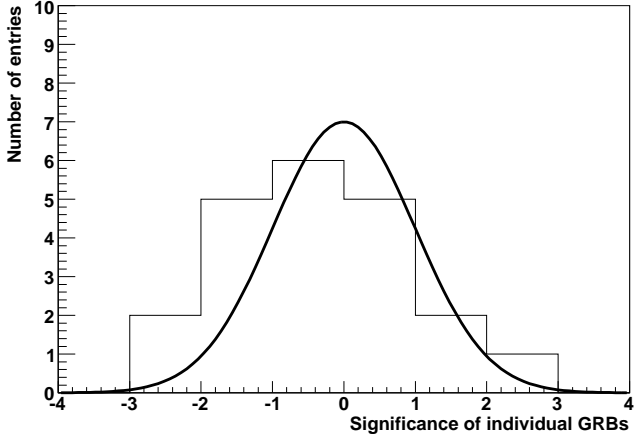}
      \caption{Distribution of the statistical significance (\emph{histogram}) as derived from the observations of 20 GRBs using soft-cut analysis. The mean is $-$0.4 and the standard deviation is 1.4. Each entry corresponds to one GRB. The \emph{solid line} is a Gaussian function with mean zero and standard deviation unity.}
         \label{soft_sigma_dist}
   \end{figure}

No evidence of a significant excess of VHE $\gamma$-ray events from any of the
GRB positions given in Table~\ref{GRBtable1} during the period covered by the
H.E.S.S. observations was found. The number of on-source ($N_{\rm on}$)
and off-source events ($N_{\rm off}$), normalization factor ($\alpha$),
excess, and statistical significance\footnote{calculated by eq. (17) in
  \citet{LiMa83}} of the excess in standard deviations ($\sigma$)
are given for each of the 21 GRBs in
Table~\ref{GRB_stat}. The results for GRB~030821 are given in
Sect.~\ref{sect_030821}. Figure~\ref{soft_sigma_dist} shows the distribution of
the significance obtained from the soft-cut analysis of the
observations of each of the 21 GRBs. A Gaussian distribution with mean zero
and standard deviation one, which is expected in the case of no
detection, is shown for comparison. The distribution of the statistical
significance is consistent with this Gaussian distribution. Thus no
significant signal was found from any of the individual GRBs. A search for serendipitous source discoveries in
the H.E.S.S. FoV during observations of the GRBs also resulted in no significant
detection. The 99.9\% confidence level (c.l.) flux upper limits (above $E_\mathrm{th}$) have been calculated using the method
of \citet{feldman98} for both standard cuts (assuming
$\Gamma=2.6$) and soft cuts (assuming $\Gamma=5$), and are included in
Table~\ref{GRB_stat}. The limits are as observed on Earth, i.e. the
EBL absorption factor was not taken into account. The systematic error on a
H.E.S.S. integral flux measurement is estimated to be $\sim$20\%, and it was not
included in the calculation of the upper limits.

For those GRBs with reported
redshifts, the effect of the EBL on the H.E.S.S. limits can be
estimated. Using the EBL model P0.45 described in~\citet{Aha06_EBL_nature}, differential upper
limits (again assuming $\Gamma=5$) at the energy threshold were calculated from the integral
upper limits obtained using soft-cut analysis. These upper limits, as
well as those calculated without taking the EBL into account, are shown in Table~\ref{GRB_differential_flux_ULs}.

\begin{landscape}
\begin{table}
\begin{minipage}[t]{\columnwidth}
\caption{H.E.S.S. observations of GRBs from March 2003 to October
  2007.}
\label{GRB_stat}
\centering
\renewcommand{\footnoterule}{}  
\begin{tabular}{l@{}rrc@{ }c|ccl@{}c@{}c@{ }rc|ccl@{}c@{}c@{ }rc@{  }cl}
\hline\hline
            &               &         &    &      &
\multicolumn{7}{|c|}{Standard-cut analysis} &  \multicolumn{7}{c}{Soft-cut
  analysis} & \multicolumn{2}{c}{Temporal analysis} \\
     GRB\footnote{The GRBs are listed in the order of the ranking scheme described in
  Sect.~\ref{sect_rank}. GRB~030821 is not listed, the results of which are given in Sect.~\ref{sect_030821}. }
     & $T_{\rm start}$ & Exposure & $N_{\rm tel}$ & Z.A.   & $N_\mathrm{ON}$ & $N_\mathrm{OFF}$ & $\alpha$ & Excess & Signi- & $E_{\rm th}$ & Flux ULs & $N_\mathrm{ON}$ & $N_\mathrm{OFF}$ & $\alpha$ & Excess & Signi- & $E_{\rm th}$ & Flux ULs & $\chi^2$/d.o.f. & P($\chi^2$) \\
            & (min)           & (min)     &               & ($\degr$) &                 &                  &          & & ficance   & (GeV)            & (cm$^{-2}$~s$^{-1}$) & &              &          &  & ficance & (GeV)            & (cm$^{-2}$~s$^{-1}$) \\
    \hline
    070621  & 6.5           & 234.6    & 4 & 16      & 204 & 2273 & 0.091 &
    -2.6 & -0.18 &  250  &  $2.8\times10^{-12}$  & 731 & 5903 & 0.13 & -6.9 &
    -0.24 & 190 & $5.6\times10^{-12}$ & 19.2/28 & 0.89 \\
    050801  & 15.0          & 28.2     & 4 & 43      & 13 & 173 & 0.091 & -2.7
    & -0.68 & 400 & $3.2\times10^{-12}$ & 46 & 442 & 0.13 & -9.3 & -1.2 & 310
    & $1.6\times10^{-11}$ & 0.168/3 & 0.98 \\
    070429A & 64            & 28.2     & 4 & 23      & 4 & 78 & 0.091 & -3.1 &
    -1.2 & 290 & $2.4\times10^{-12}$  & 20 & 203 & 0.13 & -5.4 & -1.0  & 220 &
    $1.0\times10^{-11}$ & 6.39/3 & 0.094  \\
    \multirow{2}{*}{041211B$^{a}$~$\Big\{$}  & 567.1         & 14.2     & 3 & 64      & 9 & 87 & 0.11 & -0.67 &
    -0.21 & 1850 & $6.8\times10^{-12}$ & 27 & 236 & 0.17 & -12 & -1.9 & 1360 &
    $2.6\times10^{-11}$ & \multirow{2}{*}{$\Big\}$~~14.6/14} & \multirow{2}{*}{0.40} \\
            & 742.3         & 112.3    & 4 & 44      & 76 & 1247 & 0.063 & -1.9 & -0.21 & 380 & $3.7\times10^{-12}$ & 317 & 4353 & 0.083 & -46 & -2.4 & 280 & $1.8\times10^{-11}$ & & \\
    \multirow{2}{*}{071003\footnote{Three- and four-telescope data are presented.}~~~$\Big\{$}  & 623.3         & 56.2     & 4 & 35      & 16 & 272 & 0.10 & -11 &
    -2.2 & 390 & $1.0\times10^{-12}$ & 97 & 785 & 0.14 & -15 & -1.4 & 280 &
    $1.4\times10^{-11}$ & \multirow{2}{*}{$\Big\}$~~32.3/12} & \multirow{2}{*}{0.0012} \\
            & 691.1         & 56.2     & 3 & 41      & 25 & 204 & 0.10 & 4.6 & 0.93 & 480 & $5.6\times10^{-12}$ & 79 & 547 & 0.14 & 0.86 & 0.091 & 340  & $1.5\times10^{-11}$ & & \\
    041006  & 626.1         & 81.9     & 4 & 27      & 80 & 770 & 0.10 & 3 &
    0.32 & 200 & $1.1\times10^{-11}$ & 302 & 1974 & 0.14 & 20 & 1.1 & 150 &
    $6.8\times10^{-11}$ & 8.89/9 & 0.45 \\
    070419B & 907           & 56.4     & 4 & 47      & 28 & 391 & 0.091 & -7.5
    & -1.3 & 700 & $2.4\times10^{-12}$ & 121 & 1069 & 0.13 & -13 & -1.0 & 520
    & $7.5\times10^{-12}$ & 11.9/6 & 0.064 \\
    060526  & 284.2         & 112.8    & 4 & 25      & 93 & 1068 & 0.10 &
    -13.8 & -1.3  & 280 & $2.9\times10^{-12}$ & 492 & 3711 & 0.14 & -38 & -1.6
    & 220 & $9.2\times10^{-12}$ & 19.8/12 & 0.072 \\
    070808  & 306.2        & 112.8     & 4 & 34      & 49 & 659 & 0.091 & -11
    & -1.4 & 310 & $3.2\times10^{-12}$  & 209 & 1733 & 0.13 & -7.6 & -0.49 &
    260 & $7.5\times10^{-12}$ & 15.8/12 & 0.20 \\
    070721B & 925.7         & 103.8    & 4 & 40      & 59 & 984 & 0.063 & -2.5
    & -0.31 & 440 & $1.4\times10^{-12}$ & 237 & 2676 & 0.083 & 14 & 0.89 & 320
    & $8.8\times10^{-12}$ & 15.5/11 & 0.16 \\
    061110A & 407.68        & 112.8    & 4 & 25      & 76 & 838 & 0.093 & -1.9
    & -0.21 & 280 & $4.3\times10^{-12}$ & 314 & 2671 & 0.13 & -20 & -1.0 & 200
    & $8.4\times10^{-12}$ & 4.66/11 & 0.95 \\
    030329\footnote{A slightly different analysis technique was used, see Sect.~\ref{sample_analysis}. Soft-cut analysis is not available for this observation.}  & 16493.5       & 28.0     & 2 & 60      & 4 & 26 & 0.14 & 0.27 &
    0.13 & 1360  & $2.6\times10^{-12}$ & & & & & $\cdots$ & & & 5.93/3 & 0.12 \\
    050726  & 772.7         & 112.8    & 4 & 40      & 107 & 1031 & 0.083 & 21
    & 2.1 & 320 & $7.1\times10^{-12}$  & 333 & 2619 & 0.11 & 42 & 2.3 & 260 &
    $3.4\times10^{-11}$ & 14.7/12 & 0.26\\
    050209  & 1208.5        & 168.6    & 4 & 48      & 104 & 1096 & 0.11 & -18
    & -1.6 & 480 & $4.4\times10^{-12}$ & 528 & 4204 & 0.14 & -73 & -2.8 & 340
    & $1.5\times10^{-11}$ & 36.3/18 & 0.0065 \\
    070612B & 901.7         & 112.8    & 4 & 18      & 104 & 1190 & 0.091 &
    -4.2 & -0.39 & 240 & $4.1\times10^{-12}$ &  415 & 3233 & 0.13 & 11 & 0.51
    & 180 & $1.5\times10^{-11}$ & 4.87/12 & 0.96 \\
    060403  & 820.4         & 52.8     & 4 & 39      & 33 & 252 & 0.091 & 10 &
    1.9 & 440 & $4.8\times10^{-12}$ &  128 & 875 & 0.13 & 19 & 1.6 & 320 &
    $1.3\times10^{-11}$ & 10.4/6 & 0.11 \\
    060505  & 1163          & 111      & 4 & 42      & 99 & 837 & 0.091 & 23 &
    2.4 & 520 & $5.6\times10^{-12}$ & 339 & 2740 & 0.13 & -3.5 & -0.18 & 400 &
    $3.9\times10^{-12}$ & 22.1/12 & 0.036 \\
    050509C & 1289          & 28.2     & 4 & 22      & 31 & 344 & 0.083 & 2.3
    & 0.41 & 200 & $1.7\times10^{-11}$ & 112 & 965 & 0.11 & 4.8 & 0.43 & 150 &
    $1.5\times10^{-10}$ & 0.301/3 & 0.96 \\
    070721A & 893.5         & 112.8    & 4 & 30      & 90 & 1436 & 0.059 & 5.5
    & 0.58 & 320 & $6.5\times10^{-12}$ & 280 & 3837 & 0.077 & -15 & -0.86 &
    260 & $1.3\times10^{-11}$ & 6.78/12 & 0.87 \\
    070724A & 927.5         & 84.6     & 4 & 23      & 73 & 720 & 0.091 & 7.5
    & 0.88 & 260 & $7.3\times10^{-12}$ & 246 & 2042 & 0.13 & -9.3 & -0.55 &
    200 & $1.0\times10^{-11}$ & 14.3/9 & 0.11 \\
    070209  & 926.7         & 56.4     & 4 & 41      & 37 & 444 & 0.091 & -3.4
    & -0.51 &  480 & $2.3\times10^{-12}$  & 185 & 1442 & 0.13 & 4.8 & 0.33 &
    370 & $1.1\times10^{-11}$ & 5.35/6 & 0.50 \\
    \hline
\end{tabular}
\end{minipage}
\end{table}
\end{landscape}


\begin{table}
\begin{minipage}[t]{\columnwidth}
\caption{Differential flux upper limits at the energy thresholds from the H.E.S.S.
  observations of GRBs with reported redshifts.}  
\label{GRB_differential_flux_ULs}
\centering
\renewcommand{\footnoterule}{}  
\begin{tabular}{llrcc}
  \hline\hline
  GRB   & Redshift & $E_\mathrm{th}$ (GeV) & $F_\mathrm{UL}$\footnote{Limits are given in units of $\mathrm{cm}^{-2}\,\mathrm{s}^{-1}\,\mathrm{GeV}^{-1}$.} & $F_\mathrm{corrected}$$^a$ \\
  \hline
  060505  & 0.0889  & 400 & 3.9$\times10^{-14}$ & 5.8$\times10^{-14}$  \\
  030329  & 0.1687  & 1360 & 7.6$\times10^{-15}$ & 9.7$\times10^{-14}$  \\
  070209  & 0.314  & 370 & 1.2$\times10^{-13}$ & 8.7$\times10^{-13}$  \\
  070724A  & 0.457  & 200 & 2.1$\times10^{-13}$ & 1.0$\times10^{-12}$  \\
  041006  & 0.716  & 150 & 1.8$\times10^{-12}$ & 2.7$\times10^{-11}$  \\
  061110A & 0.758  & 200 & 1.7$\times10^{-13}$ & 1.7$\times10^{-11}$  \\
  050801  & 1.56  & 310 & 2.1$\times10^{-13}$ & \footnote{The limits corrected for EBL absorption are $>$10 orders of magnitude larger than that observed.}  \\
  071003\footnote{Only 4-telescope data were used.}
          & 1.604  & 280 & 2.0$\times10^{-13}$ & $^b$ \\
  060526  & 3.21  & 220 & 1.7$\times10^{-13}$ & $^b$ \\
  070721B  & 3.626  & 320 & 1.1$\times10^{-13}$ & $^b$ \\
  \hline
\end{tabular}
\end{minipage}
\end{table}

\begin{table}
\caption{Combined significance of 3 subsets of GRBs selected based on the
  requirements listed in Sect.~\ref{sect_stacking}}  
\label{subset}
\centering
\begin{tabular}{lcrr}
\hline\hline
                & Number  & Soft-cut & Standard-cut \\
                & of GRBs & analysis & analysis \\
\hline
Sample A        & 10      &  -2.13 & -1.81 \\
Sample B        & 6       &  -0.20 & 1.45  \\
Sample C        & 11      &  -0.53 & 0.48  \\
all GRBs        & 21      &  -1.98 & -0.18 \\
\hline
\end{tabular}
\end{table}

\subsection{Stacking analysis}
\label{sect_stacking}
Although no significant excess was found from any individual GRB, co-adding
the excess events from the observations of a number of GRBs may reveal a
signal that is too weak to be seen in the data from one GRB, provided that
the PSFs of the H.E.S.S. observations are bigger than the error box of the
GRB positions (which is the case, see Sect.~\ref{sample_analysis}). Firstly, stacking
of all GRBs (except GRB~030821, which has a high positional uncertainty) in the sample was performed. This yielded a total
of $-$157 excess events and a statistical significance of $-$1.98 using the
soft-cut analysis. Use of standard cuts produced a similar result (see
Table~\ref{subset}). Secondly, combining the significance of the results from three selected
subsets extracted from the whole sample was
performed. The \emph{a priori} selection criteria were to choose those GRBs
with a higher expected VHE flux or a lower level of EBL absorption. The
following requirements were used to select three subsets:

\begin{description}
 \item[Sample A:] the first 10 in the ranking described in Sect.
   \ref{sect_rank};
 \item[Sample B:] all GRBs with a measured redshift $z<1$;
 \item[Sample C:] all GRBs with a soft-cut energy threshold lower than 300~GeV
   and with either a measured redshift $z<1$ or with an unknown redshift.
\end{description}

The result is shown in Table~\ref{subset}. As can be seen, there is no
significant evidence of emission in any of these subsets.

\subsection{Temporal analysis}
As possible VHE radiation from GRBs is expected to vary with time, a temporal analysis to search for deviation from zero excess in the observed data was performed. Soft-cut analysis was used for all GRBs (except GRB~030329)
since this analysis has a lower energy
threshold and a better acceptance of $\gamma$-rays and cosmic rays and therefore
increases the statistics. The $\gamma$-like excess events were binned in 10-minute time intervals
for each GRB data set and were compared to the assumption of no excess
throughout the observed period. 
The $\chi^2$/d.o.f. value and the
corresponding probability are shown in Table~\ref{GRB_stat} for each
GRB. Within the whole sample, the lowest probability that the
hypothesis that the excess was zero throughout the observation period is
correct is $1.2\times10^{-3}$ (for GRB~071003) and no significant
deviation from zero within any of the GRB temporal
data was found. Standard-cut analysis produced consistent results.

\subsection{GRB~070621: Observations of a GRB with the fastest reaction and the longest
  exposure time}
\label{sect_070621}
\begin{figure}
   \centering
   \includegraphics[width=9.0cm]{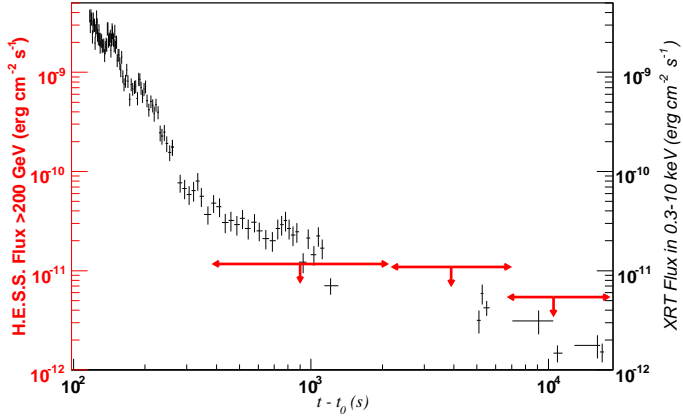}
      \caption{The 99.9\% confidence level energy flux upper limits (in red) at energies
        $>$200 GeV derived from H.E.S.S. observations at the position of
        GRB~070621. The ends of the horizontal lines indicate the start
        and end times of the observations from which the upper limits
        were derived. The XRT energy flux in the 0.3--10 keV band is shown in
        black for comparison~\citep{evans07}.}
         \label{GRB070621_lc}
   \end{figure}

GRB~070621 is the highest-ranked GRB in the sample
(Sect.~\ref{sect_rank}), i.e. it has the highest relative expected VHE flux at
the start time of the observations. The duration of the \emph{Swift} burst was
$T_{90}\sim33$s, thus clearly classifying the burst as a long GRB. The fluence in
the 15--150 keV band was $\sim$4.3$\times10^{-6}$\ erg\ cm$^{-2}$. The XRT
light curve is represented by an initial rapidly-decaying phase and a shallow phase, with the transition happening around $t_0 + 380$s where $t_0$ denotes the trigger
time~\citep{sbarufatti07}. Despite extensive optical monitoring, no fading
optical counterpart was found. The H.E.S.S. observations started at $t_0+420$s
and lasted for $\sim$5 hours, largely coincident with the X-ray shallow
phase. These observations were both the most prompt and the longest among those presented. Figure~\ref{GRB070621_lc} shows the 99.9\% H.E.S.S. energy flux upper limits
above $200$~GeV (using soft-cut analysis), together with
the XRT results \citep{evans07}. As seen, the limits for
this period are at levels comparable to the X-ray energy flux during the same
period. Unfortunately the lack of redshift information for
this burst prevents further interpretation of the limits.

\subsection{GRB~030821: Observations of a GRB with a high positional
  uncertainty}
\label{sect_030821}
Some GRBs, such as GRB~030821, have a high uncertainty in position; with a relatively large camera
FoV ($\sim$5$\degr$), the H.E.S.S. telescopes are able to cover the whole
positional error box of such GRBs.

\begin{figure}
   \centering
   \includegraphics[bb= 14 14 582 387,clip,width=9.0cm]{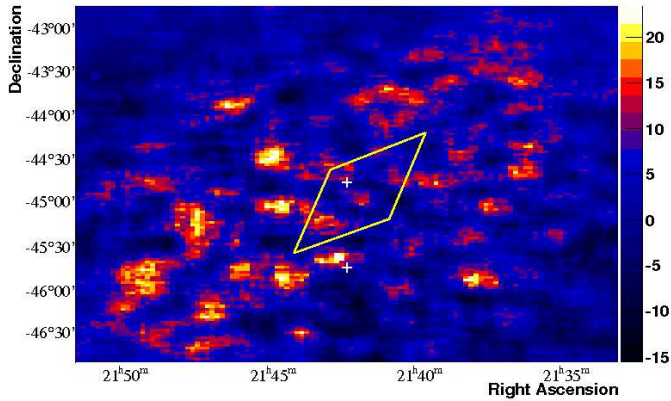}
     \caption{The $\gamma$-like excess events in the region of the
        GRB~030821. The error box shows the position of the burst localized by
        IPN triangulation \citep{hurley2359}. The colour (grey) scale is set such
        that the blue/red (black/grey) transition occurs at the $\sim$1.5$\sigma$
        significance level. The sky map was derived using two observations
        pointing at two different positions (marked by crosses), resulting in a non-uniform distribution of events in the map.}
         \label{030821overlay}
   \end{figure}

Observations of GRB~030821 started 18 hours after the burst
and lasted for a live-time of 55.5 minutes, with a mean Z.A. of
28$\degr$. The observations were taken when the array was under construction and only two telescopes were operating, resulting in an energy threshold of 260~GeV. The GRB has a relatively high uncertainty in
position as determined from \emph{IPN} (the third Interplanetary Network)
triangulation \citep{hurley2359}, and its error box is bigger than the
PSF of H.E.S.S. However, because of the relatively large FoV of the camera, the whole
error box, and thus the possible GRB position, is within the H.E.S.S. FoV. The sky excess map overlaid with the error box is shown in
Fig.~\ref{030821overlay}. As can be seen, there is no significant excess at any position within the error
box. The sky region with the largest number of peak excess events is located in the south-eastern part of the error box. Using a point-source analysis centred at this peak, a flux upper limit (above 260~GeV) of
$\sim$1.7$\times10^{-11}$~cm$^{-2}$~s$^{-1}$ was derived. Since an upper limit derived for any location in the error box with fewer excess events is \emph{lower} than this value\footnote{A larger excess implies a higher value of the upper limit, since the integrated exposure, that depends on Z.A. and $\theta_\mathrm{offset}$ of the observations, is largely the same over the whole error box.}, it may be regarded as a \emph{conservative} upper limit of the VHE flux associated with GRB~030821 during the period of the H.E.S.S. observations.


\section{Discussion}

The upper limits presented in this paper are among the most stringent
ever derived from VHE $\gamma$-ray observations of GRBs during the
afterglow period. In fact, the
99.9\% confidence level limits (in energy flux) are at levels comparable to the X-ray energy flux as observed by \emph{Swift}/XRT
during the same period (see, e.g. Fig.~\ref{GRB070621_lc}).
Unless most of the GRBs are located at high redshifts and thus their VHE flux is severely
absorbed by the EBL (this possibility is discussed below), one expects detection of the predicted VHE component with
energy flux levels comparable to those in X-rays in some
scenarios~\citep{dermer00,wang01,zhang01,peer05,fan08}.

On the other hand, the unknown redshifts of many of the GRBs in the sample
(including GRB~070621, the highest-ranking, which is discussed in Sect.~\ref{sect_rank}) complicate the physical interpretation of the data,
because EBL absorption at VHE energies is severe for a GRB with $z>1$. The
mean and median redshift of the 10 GRBs with reported redshifts is 1.3 and
0.7, respectively. If the 12 GRBs without redshift have the same redshift distribution, one would expect $\sim$40\% of them ($\sim$5~GRBs) to have $z<0.5$. In this case, the EBL absorption may not preclude the detection of the predicted VHE $\gamma$-rays for the GRB sample
presented here\footnote{The optical depth of EBL absorption for a $\sim$100~GeV photon
is $\sim$3 at $z=1$, according to the P0.45 model demonstrated in~\citet{Aha06_EBL_nature}.}.

There is no reported X-ray flare during the H.E.S.S. observational time windows, therefore no conclusion on whether or not X-ray flares are accompanied by VHE flares, as well as the origin of X-ray flares, can be drawn. If UHECRs are generated in nearby GRB sources, as suggested by some authors, a detectable VHE flux is expected from nearby GRBs. Therefore, although the unknown redshifts of a significant fraction of GRBs in our sample (12 out of 22) and the uncertainty in the modeled VHE temporal evolution are surely in play, the results presented here do not indicate (but also not exclude) that GRBs are dominant sources of UHECRs.

\section{Outlook}

The data from our sample of 22 GRBs do not provide any evidence for a strong VHE $\gamma$-ray
component from GRBs during the afterglow phase. EBL absorption can explain the lack of detection in our sample. 
However, this does not exclude a
population of GRBs that exhibit a strong VHE component. While the EGRET
experiment did not detect MeV--GeV photons from most BATSE GRBs in its FoV, some strong bursts
(e.g. GRB~940217) have proved to emit delayed emission, $\sim$1.5
hours after the burst, at energies as high as $\sim$20~GeV~\citep{hurley94}. With \emph{Fermi}'s observations of GRBs having started in mid-2008, it is likely that our knowledge of the high-energy emission of GRBs will be improved in the near future.

The future prospects for detection at VHE energies rely on the likelihood of observing a GRB with low redshift (e.g. $z<0.5$) early enough. In the cases where there is no detection, sensitive and early upper limits on the intrinsic VHE luminosity of these nearby GRBs will still improve our understanding of the radiation mechanisms of GRBs. The existence of a distinct population of low-luminosity (LL) GRBs was suggested based on the high detection rate of low-redshift LL GRBs such as GRB~980425 and GRB~060218~\citep[e.g.][]{Soderberg04_nat}. Due to their proximity, they are good targets for VHE observations. Since they are sub-energetic compared to other GRBs, they may be accompanied by a lower VHE luminosity. On the other hand, if most radiation are emitted at high energies, the detection probability would be much higher.



\citet{Franceschini08} claimed a very small $\gamma$-ray opacity due to EBL absorption. The optical depth is about a factor of three less than the one we used, depending on the energies~\citep{Aha06_EBL_nature}. Therefore, on-going GRB observations with H.E.S.S., as well as other ground-based VHE detectors, are crucial to test this model.

\section{Conclusions}

During 5 years of operation (2003--2007), 32 GRBs were
observed during the afterglow phase using the H.E.S.S. experiment. Those 22 GRBs with high-quality data were
analysed and the results presented in this paper. Depending on the visibility and observing
conditions, the start time of the observations varied from minutes to hours
after the burst.

There is no evidence of VHE emission from any individual
GRB during the period covered by the H.E.S.S. observations, nor from stacking
analysis using the whole sample and \emph{a priori} selected sub-sets of GRBs. Fine-binned temporal data revealed no short-term variability from any observation and no indication of VHE signal from any of these time bins was found. Upper limits of VHE $\gamma$-ray flux during the observations from the GRBs were
derived. These 99.9\% confidence level energy flux upper limits are
at levels comparable to the contemporary X-ray energy flux. For those GRBs with reported redshifts,
differential upper limits at the energy threshold after correcting for EBL absorption
are presented.

H.E.S.S. phase II will have an energy threshold of about 30~GeV. With much less
absorption by the EBL at such low energies, it is hoped that the H.E.S.S. experiment will
enable the detection of VHE $\gamma$-ray counterparts of
GRBs.

\begin{acknowledgements}

The support of the Namibian authorities and of the University of Namibia
in facilitating the construction and operation of H.E.S.S. is gratefully
acknowledged, as is the support by the German Ministry for Education and
Research (BMBF), the Max Planck Society, the French Ministry for Research,
the CNRS-IN2P3 and the Astroparticle Interdisciplinary Programme of the
CNRS, the U.K. Science and Technology Facilities Council (STFC),
the IPNP of the Charles University, the Polish Ministry of Science and
Higher Education, the South African Department of
Science and Technology and National Research Foundation, and by the
University of Namibia. We appreciate the excellent work of the technical
support staff in Berlin, Durham, Hamburg, Heidelberg, Palaiseau, Paris,
Saclay, and in Namibia in the construction and operation of the
equipment. P.H. Tam acknowledges support from
IMPRS-HD. This work has made use of the GCN Notices and Circulars provided
by NASA's Goddard Space Flight Center, as well as data supplied by
the UK \emph{Swift} Science Data Centre at the University of Leicester.

\end{acknowledgements}

\end{document}